Phase resetting reveals network dynamics underlying a bacterial cell cycle

Yihan Lin[1, 4], Ying Li[2, 4], Sean Crosson[3], Aaron R. Dinner[1, 4, *], and Norbert F. Scherer[1, 4, *]

Department of Chemistry[1], Physics[2], and Biochemistry and Molecular Biology[3], and Institute for Biophysical Dynamics[4], University of Chicago, 929 East 57th Street, Chicago, IL 60637

*E-mail: dinner@uchicago.edu; nfschere@uchicago.edu

Abstract

Genomic and proteomic methods yield networks of biological regulatory interactions but do not provide direct insight into how those interactions are organized into functional modules, or how information flows from one module to another. In this work we introduce an approach that provides this complementary information and apply it to the bacterium Caulobacter crescentus, a paradigm for cell-cycle control. Operationally, we use an inducible promoter to express the essential transcriptional regulatory gene ctrA in a periodic, pulsed fashion. This chemical perturbation causes the population of cells to divide synchronously, and we use the resulting advance or delay of the division times of single cells to construct a phase resetting curve. We find that delay is strongly favored over advance. This finding is surprising since it does not follow from the temporal expression profile of CtrA and, in turn, simulations of existing network models. We propose a phenomenological model that suggests that the cell-cycle network comprises two distinct functional modules that oscillate autonomously and couple in a highly asymmetric fashion. These features collectively provide a new mechanism for tight temporal control of the cell cycle in C. crescentus. We discuss how the procedure can serve as the basis for a general approach for probing network dynamics, which we term chemical perturbation spectroscopy (CPS).

# Author Summary

During the cell cycle, the cell progresses through a series of stages that are associated with various cell cycle events such as replication of genetic materials. Genetic and molecular dissections have revealed that the cell cycle is regulated by a network of interacting molecules that produces oscillatory dynamics. The major cell cycle regulators have been identified previously in different species and the activity of these regulators oscillates. However, the question of how cell cycle regulators coordinate different cell cycle events during the cell cycle remains controversial. Here, we investigate this question in a model bacterial system for cell cycle, Caulobacter crescentus. We perturb the expression of the master cell cycle regulator ctrA in a pulsatile fashion and quantify the response of the cell cycle to such perturbations. The measured response is contradictory to the existing mechanism of Caulobacter cell cycle control, which views the cell cycle progression as a sequential activation/inhibition process. We propose a new model that involves coupling of multiple oscillators and show the quantitative agreement between this new model and our measurements. We expect this procedure to be generalized and applied to a broad range of systems to obtain information that complements that obtained from other methods.

The regulatory network that coordinates oscillating periods of growth, chromosome replication, and division is among the most important in a cell[1]. It is emerging that the cell cycle network, like others, is organized into functional modules [1-5]. Each module is sequentially activated or inhibited by key cell cycle regulatory proteins, whose concentrations oscillate with the same period to ensure irreversibility and a "once-per-cell-cycle" occurrence of each process [1,2]. However, in both prokaryotes and eukaryotes there is increasing evidence that internal regulatory modules (i.e., a set of chemical reactions associated with a key sub-function of the overall cell cycle) can run autonomously. For example, in the bacterium C. crescentus several rounds of chromosome replication can occur under conditions where activity of the master cell cycle regulator CtrA is largely suppressed[6], and certain C. crescentus mutants can undergo multiple cell constrictions within one cell cycle [7-9]. In budding yeast, cell cycle modules such as budding [10], transcription [11], centrosome replication [12], and Cdc14 localization [4,13] can run independently of Cdk activity. This raises the question of how individual modules interact to generate robust sequences of events.

The interactions defining the connectivity of a regulatory network, such as that controlling the cell cycle, can be dissected in a traditional manner by functional reconstitution [14]. However, this does not provide information about the integrated dynamics of the interacting network as a whole. Alternatively, by applying appropriate perturbations to an intact network, one can determine the dynamics of the response of one or more measureable parameters and infer global properties of the network that underlie a given process. We refer to this as probing the topology of the functional relations of the network. Such an approach is analogous to circuit analysis in electrical engineering and time-resolved spectroscopies employed in chemistry and physics [15,16]. Here we report a periodic perturbation approach that provides insight into the systems-level control features of a bacterial cell cycle.

Specifically, we study Caulobacter crescentus because its cell cycle regulatory network has been well-characterized both genetically and biochemically [17] and quantitative models have been reported [18-20]. The life cycle of C. crescentus begins as a non-reproductive motile swarmer cell, with chromosome replication inhibited by the cell cycle master regulator CtrA [21] binding to the replication origin [6]. The C. crescentus swarmer cell then differentiates into a reproductive sessile stalked cell (i.e., the mother); this cell differentiation event is concomitant with proteolytic clearing of CtrA from the cell. The stalked cell then commences DNA replication, cell growth, FtsZ ring formation, and membrane fission to yield a daughter swarmer cell and regenerate the mother stalked cell [22] (Fig. 1A left). While swarmer progeny remain in a gap-like phase prior to differentiation, the stalked progeny can continue to reproduce for tens of generations [20,23]. Thus the stalked cell behaves as a self-sustained oscillator. The CtrA concentration profile during the stalked cell cycle, shown schematically in Fig. 1A, is low in the early stalked cell, reaches a maximum at the late-predivisional stage, then decreases rapidly in the stalked compartment (post-constriction) prior to initiation of a successive round of reproduction [17,24,25]

The design of our experiment is as follows (Fig. 1B). In contrast to knockout experiments that completely eliminate an element of a regulatory network, we seek to quantitatively perturb the expression of a molecule and analyze the resulting change in system dynamics. To this end, we engineer a mutant strain that lacks ctrA and then introduce a xylose-inducible ctrA on a plasmid [26]. It is important to note that this strain (with the plasmid) grows and divides in an essentially normal fashion in the presence of constant xylose concentration ($0.9 \times 10^{-4}$%, w/v; Fig. S1). This is made possible by the fact that the active form of CtrA is the phosphorylated protein [24,25], which can still oscillate even though the gene is transcribed at a constant rate. We then use a microfluidic device to toggle between low and high levels of xylose. Although the inducer concentrations are such that the protein should always be present at levels that permit division, the periodic pulses of expression must indirectly increase the amount of phosphorylated protein because they cause division to synchronize. We measure division times of single cells and use them to determine the advance or delay of the cell cycle as a function of its phase when a pulse arrives. This response defines the phase resetting curve (PRC), which informs a mathematical analysis that reveals two important insights into the cell-cycle network: (i) it comprises functional modules that oscillate autonomously and (ii) the coupling between these units is highly asymmetric such that CtrA acts to brake rather than drive the cell cycle. We validate this model by quantitative comparison with independent experimental data. We discuss molecular mechanisms for realizing the elucidated functional features and their potential biological advantages.

## Results/Discussion

**Validation of the experimental construct.** As discussed in the Introduction, we examine the quantitative change in division times in response to pulses in CtrA in a C. crescentus ctrA mutant strain with a xylose-inducible copy of ctrA. Unless otherwise indicated, we switch between xylose levels of $0.9 \times 10^{-4}$% (w/v) and 0.03%. To ensure that CtrA was not limiting, we first measured inter-division times for fixed xylose levels. The strain is viable for xylose concentrations $0.9 \times 10^{-4}$%. The mean inter-division time at $0.9 \times 10^{-4}$% xylose is 68.1 ± 15.6 min (N=5160 cell division events; temperature = $32.0^{\circ}$C), which is comparable to the wild-type, although it should be noted that the noise (standard deviation/mean) is larger (Fig. S1).

In the periodic experiment, we stimulate a population of surface attached stalked cells cultured in a Y-shaped microfluidic device [20] with a pulse train that alternates between low and high xylose concentrations (Fig. 1B left, Methods). We explored a range of external pulse periods that was centered about the mean intrinsic cell-cycle time (i.e., 68.1 min). Fig. 2 shows the results obtained with pulses of 15 min high xylose (0.03%) and 50 min low xylose concentration ($0.9 \times 10^{-4}$%) (i.e., external oscillator period is 65 min). The entrainment of the cell cycle to this external periodic pulse train can be readily visualized in the growth curve that is constructed from the measured single cell divisions, only counting progeny of the original stalked cells (Fig. 2A upper panel; Methods). The linear growth of the initial portion of the curve is due to asynchronous division, and the subsequent stepwise growth corresponds to synchronous division [27]. Similar synchronization was recently realized for synthetic genetic oscillators [28]. Here, the synchronization confirms that the pulses in (initially non-phosphorylated and thus inactive) CtrA are sufficient to perturb the cell cycle and serve as the basis for phase resetting. This observation is consistent with the idea that the active form of CtrA follows the overall CtrA protein level closely owing to rapid phosphorylation during the stalked cell cycle [24,25,29,30].

The phase resetting curve is sufficient to capture the statistics of division. Many (~20) divisions were followed for each cell. Each division event is indicated by a dot in Fig. 2A (lower panel), with the timing of the event on the horizontal axis and its lineage on the vertical axis (labeling each original stalked cell by its cell-cycle phase immediately prior to the first pulse). By construction, the initial condition is a diagonal line in this representation; it reflects an asynchronous population with a large dispersion in phase over the cell cycle immediately before perturbation. The dispersion from a line for divisions at negative time in this reference frame (i.e., reading to the left in Fig. 2A) shows the intrinsic noise in the cell cycle. After the pulse train starts (positive time), the population gradually evolves to a distribution about a vertical line, indicating synchronous division. The distribution narrows and the synchronization, as quantified by an information theoretic measure, the synchronization index [31] (see Methods), increases as more pulses are delivered (Figs. 2B and 2C). Thus quantitative metrics support phase-locking of the cell cycle.

The phase-locking efficacy varies across the range of external pulse frequencies (i.e., inverse periods) explored (Fig. 3A and Fig. S5) owing to the intrinsic noise in the cell cycle. Stronger entrainment occurs when ctrA is induced for 15 min than 10 min for the same overall pulse frequency. The synchronization index peaks when the pulse frequency equals the intrinsic cell-cycle frequency and decays more rapidly at higher frequencies than at lower frequencies. The asymmetry of the response can be seen much more clearly by using the single-cell data to construct a phase resetting curve (PRC) [32]. The PRC is the deviation of the division time for each event from the unperturbed cell cycle period plotted as a function of the start time of the pulse (relative to the previous division; Fig. 3B; Fig. S2). In constructing Fig. 3B, we assume that the responses to successive pulses are independent and pool their phase shifts. This assumption is confirmed by our data (Fig. S3) and is also justified by the fast turnover rate of CtrA [24].

Return map analysis (Fig. S4) reveals that stable phase-locking corresponds to portions of PRCs with slopes between 0 and –2 (see Eqs. (S1)-(S6) and Figs. S5). The experimental PRCs (Fig. 3B) are further employed in Langevin equation simulations to recapitulate the measured synchronization responses of cell population to external periodic (Fig. S6-7) and non-periodic (Fig. S8) pulse trains. These results together indicate that the constructed PRCs, which are average responses, are sufficient to capture the division statistics of interest.

The cell cycle is more readily delayed than advanced. Consistent with the frequency response above, our experimental PRCs demonstrate that the cell cycle response to CtrA pulsation exhibits greater delay than advance. This asymmetry is our main experimental result. Our finding is surprising considering the measured CtrA temporal concentration profile, which has a slow rise and a rapid fall (~70 min and ~10 min, respectively, under conditions with a stalked cell period of ~80 min)[33]. Perturbation pulses that occur during the rise will tend to advance the CtrA oscillation, while pulses applied during the fall will tend to delay the CtrA oscillation (Fig. 3C). The slow rise and rapid fall should thus favor advance over delay. A mathematical model that is based on current molecular knowledge [19] also exhibits a much more pronounced advance than delay, regardless of the choice of parameters (Fig. S9). The behavior of the molecular model can be understood as follows. CtrA accumulates during the stalked phase and peaks at the pre-divisional stage. This accumulation positively feeds to a proteolytic system that rapidly turns over CtrA within a short time. In this way, the different modules function like gears in a machine—there is no clutch to allow variable coupling between the "engine" and "transmission", and cell division is locked to CtrA oscillation.

The CtrA-dependent PRCs that we obtained from our measurements are inconsistent with an explicit gear-like mechanism (see Fig. 3D). Corroborating this idea, the strict concentration dependence of a gear-like mechanism would predict that a decreased amplitude of the regulatory signal should either block or delay cell cycle events [4]. Indeed, in the above mathematical model [19], a reduced ctrA induction level leads to a reduced amplitude of its oscillation and a longer period. However, we showed that C. crescentus cells yield similar reproduction cycle time distributions for a wide range of constant inducer concentrations [20] (Fig. S1). Furthermore, the fact that the functional modules of the regulatory network need not all move forward at the same pace and can even run independent of the cell cycle [6-9] suggests that coupling of multiple (autonomous) oscillators is a fundamental feature of the system.

Elucidation of the form of the coupling between functional modules. Our point is not to argue for or against any particular molecular model but to show that our systems-level measurements are qualitatively inconsistent with extrapolations of behavior from the known molecular interactions. To interpret our data, in particular the PRC, we introduce a simple phenomelogical model that reveals systems-level information and can guide future studies. It comprises a core module (subscript 1 in Eq. (1)) that is coupled to a peripheral division module (subscript 2 in Eq. (1)) (Fig. 4A):

$$
\begin{aligned}
\mathrm{d}\varphi_1 / \mathrm{dt} &= \omega_1 + Z_1(\varphi_1)\Delta(t) \\
\mathrm{d}\varphi_2 / \mathrm{dt} &= \omega_2 + C(\varphi_1 - \varphi_2)
\end{aligned}
\tag{1}
$$

Here, $\varphi_1$ and $\varphi_2$ are the phases of the CtrA oscillator and the cell division oscillator respectively, $\omega_1$ and $\omega_2$ are the corresponding intrinsic phase velocities, $Z_1(\varphi_1)$ describes the response of the core module to a time-dependent perturbation $\Delta(t)$, and $C(\varphi_1 - \varphi_2)$ is the coupling from the core CtrA oscillator to the cell division oscillator and is a function of the phase difference. The function $\Delta(t)$ encodes the CtrA pulse train (i.e., it is 1 for the duration of each pulse and 0 otherwise). We take for $Z_1(\varphi_1)$ the derivative of the PRC of an existing model of the CtrA oscillator[20] (Fig. S10 and SI Text). To elucidate $C(\varphi_1 - \varphi_2)$ (Fig. 4B) we begin by noting that the perturbation never results in a stable phase difference other than the original one (see Methods). Consequently, we know that the system has a single stable point, and we can choose the zeros of $\varphi_1$ and $\varphi_2$ such that it occurs at $\varphi_1 - \varphi_2 = 0$. Mathematically, $C(0) = 0$ and $C'(0) > 0$, where the prime denotes differentiation. The slope of $C(\varphi_1 - \varphi_2)$ sets the relaxation rate; the relaxation rate in turn sets the the extent of the advance when $\varphi_1 - \varphi_2 \geq 0$ and the extent of the delay when $\varphi_1 - \varphi_2 \leq 0$. We adjusted the slopes of line segments for $\varphi_1 - \varphi_2 \geq 0$ and $\varphi_1 - \varphi_2 \leq 0$ separately to match the advance and delay observed in the 15-min pulse experiment. In this model, the phase advance of the CtrA oscillator is weakly coupled to the division oscillator, while the delay is strongly coupled (Methods).

We test the model and the elucidated coupling function by using it, without further modification, to compute the measured PRC obtained with 10-min pulses (Fig. 4C-D). This is a non-trivial test since the 10-min-pulse PRC is not a simple scalar multiple of the 15-min-pulse PRC. We see that the agreement is excellent. Crucially, the model captures the fact that the asymmetry between delay and advance is less pronounced for 10-min pulses. An additional prediction of this coupled oscillator model is that the cell cycle will become more gear-like with stronger coupling and less gear-like with weaker coupling. Indeed, weakening the coupling by lowering the amplitude of the first oscillator by decreasing the ctrA induction level reduces the coherence of the second oscillator output due to the presence of increased noise (i.e., the ratio of standard deviation of the cell inter-division time distribution over mean inter-division time increases, Fig. S1). Meanwhile, multiple cell constrictions that occur within a cell cycle [7-9] could be explained by "phase slip" between the autonomous CtrA and division modules. This asymmetrically coupled oscillator picture thus provides a theoretical foundation to explain the experimentally observed bacterial cell-cycle defects.

Molecular interpretation. What molecular components could make up the autonomous oscillator downstream of the core CtrA module? A self-sustained oscillator requires negative feedback with sufficient time delay [34,35]. Examination of the molecular details identifies the existence of an appropriate motif in the FtsZ-FtsQA interactions (Fig. S11): i) the residual transcription activity of ftsQ and ftsA from the $P_{aq}$ promoter (~25% of normal activity, estimated from [36]) in the absence of CtrA may yield sufficient expression of FtsQ and FtsA; ii) transcription of ftsA from the $P_a$ promoter is independent of CtrA [36]; iii) the time difference between FtsZ expression and Z-ring formation may provide sufficient time delay for the feedback loop; iv) the cell phase-dependent proteolytic property of FtsZ provides a negative feedback signal, i.e., the half-life of FtsZ decreases rapidly as Z-ring constriction initiates [37]. Thus Z-ring formation contributes to a time-delay while Z-ring constriction negatively regulates the stability of the division proteins. These details are encapsulated in the coupled oscillator scheme of Fig. 4A.

Broader implications. The present study is an important step beyond the recent work using simple synthetic biological oscillators [28] because we can exploit the dynamics to learn about the natural organization of the cell cycle and its design principles. Our findings are congruent with the recent observation that DnaA activity, which controls DNA replication, oscillates independently of CtrA [5], and, more generally, the "phase-locking" model proposed by Lu and Cross for budding yeast [4]. In the yeast model, the central cyclin/Cdk oscillator entrains a series of autonomous peripheral oscillators with intermittent coupling. Corroborating this picture, periodic CLN3 expression indicates that certain checkpoints in the yeast cell cycle can be abolished [38].

Given that the coupled-oscillator topology appears in the cell cycle control of multiple organisms, it is important to consider its implications and functional advantages. While the CtrA module is often viewed as the "engine" of the cell cycle, our results show that it cannot significantly accelerate division; rather, it appears to function more like a brake, slowing downstream events. This could be important for ensuring coordination of the many processes that contribute to the cell cycle. The asymmetric, diode-like, coupling function in Fig. 4B will also affect the propagation of noise from the upstream module to the downstream one. To show this, we added white noise terms to both oscillators in Eq. (1). The upstream (i.e., $\varphi_1$) noise propagates to the downstream (observed) phase through asymmetric coupling, giving rise to a skewed distribution in the (unperturbed) division times (Fig. S12). To reproduce the experimental distribution, the upstream noise needs to be ~10-fold greater than the downstream noise (Fig. S12 and SI Text). Thus, the coupled-oscillator topology filters perturbations/fluctuations that advance the CtrA oscillator phase. This reduction in noise, in turn, would prevent premature division. In addition, it would be also interesting to investigate the robustness and stability of this coupled-oscillator model through systematic non-equilibrium theoretical frameworks [39-41].

The approach that is presented here builds on the basic principle of linear response, which is central to spectroscopy and engineering. In this sense, it is a chemical perturbation spectroscopy (CPS) [42,43]. The parameterless fit of the 10-min-pulse PRC data with the model determined from the 15-min-pulse PRC establishes its suitability in the present case. Our work transcends recent linear response studies of cellular networks [44-48] by going beyond the steady-state to determine the full cell cycle response to pulsatile perturbation, as represented by the PRC. In this sense, it is most similar to [49] but we focus on extracting topological features of the regulatory dynamics rather than discriminating between specific molecular models. This analysis can be adapted to study oscillatory dynamics in other cellular systems. In the future, we envision multiple chemical perturbations, potentially with more complex waveforms, that could directly probe the bidirectional information flux between functional modules, in analogy to multi-dimensional (NMR and optical) spectroscopies[50].

## Methods

**Construction and characterization of C. crescentus ctrA mutant strains.** FC1006 was constructed by substituting the defective holdfast synthesis gene hfsA [51] in LS2535 (NA1000 ΔctrA + PxylX::ctrA) with CB15 hfsA allele [26] by double-recombination. The CB15 hfsA allele-containing plasmid was introduced to LS2535 by tri-parental mating from Top10/pNPTS 138-CB15-hfsA [26] and was selected on a 20 µg/ml kanamycin PYE plate supplemented with 0.3% xylose. Colonies were grown overnight without kanamycin selection to allow recombination, counter-selected on a sucrose containing plate, and then tested on a kanamycin plate to ensure the loss of kanamycin resistance. The successful recombinant was screened by the adhesion phenotype with the 96-well crystal violet assay [26], and confirmed by PCR amplification (MEN-SNP-70 primers, TCCCGGTCCAGTTTCAGC and  AAGTACGCGGTGGCTTCG) and restriction enzyme digestion with AvaI and BstNI. The resulting FC1006 strain has ~30% of the surface adhesion ability of wild-type CB15 after 5 hrs of induction with     0.03% xylose as characterized by the polystyrene binding assay.  The FC1071 strain was constructed by introducing the PxylX::ctrA plasmid [24] from LS2535 into FC764 (NA1000 with CB15 hfsA allele [26]).

Cell culture. Individual colonies (FC1006 or FC1071) were picked from a fresh PYE agar plate supplemented with necessary antibiotics and xylose (1 µg/ml chloramphenicol, 0.03% xylose) and grown overnight in PYE medium (1 µg/ml chloramphenicol, 0.03% xylose) in a 30°C rolled incubator. The overnight culture was diluted to $OD_{660}$=0.1 with fresh PYE (with antibiotics and xylose) and cultured for additional 2 hrs before loading into the microfluidic device with a syringe[20].

Microfluidic device and single-cell assay. Y-shaped microfluidic channels with rectangular cross-section (150µm width $\times$ 50µm height) were fabricated by rapid phototyping in poly(dimethylsiloxane) (PDMS) [52].  The PDMS and a microscope coverslip (No. 1.5) were plasma cleaned and then pressed and sealed to form Y channels with inlet and outlet ports in the PDMS. Each device contains multiple channels allowing simultaneous measurements under different conditions. Teflon tubing connectors (constructed with i.d. 0.028" /o.d. 0.048" tubing and i.d. 0.045" /o.d. 0.062" tubing) plugged with i.d. 0.012" / o.d. 0.030" tubing were connected to ports and used for solution exchange.  Before loading the bacterial cell culture, the channel was sequentially rinsed with NaOH (2M), ethanol, and autoclaved $H_2O$. After thermal equilibration inside the heated microscope enclosure and incubator, the channel was loaded with the bacterial cell culture.  Generally, ~1 hr incubation for FC1006 or ~30 min incubation for FC1071 is necessary for a sufficient number of single cells to become attached onto the glass surface of the channel. Two computer-controlled syringe pumps (PHD2000, Harvard Apparatus) that are also inside the heated (thermostated) microscope incubator were used to pump two thermally equilibrated PYE media with low and high xylose concentrations through the channel at a constant flow rate (10µL/min)[20].

Time-lapse microscopy. Time-lapse single-cell measurements were performed on an automated inverted microscope (Olympus X70) equipped with a motorized sample stage, an objective motor driver and a controller (BioPrecision stage and MAC5000 controller, Ludl Electronics). DIC microscopy was done with an Olympus UPLSAPO 100X oil objective and a light-emitting diode (LED) light source which is pulse-modulated (LEDC19 LED and LEDD1 driver, Thorlabs).  The control pulse for the LED was generated from a PCI-DAQ card (PCI-6052E, National Instrument) through a BNC adaptor interface (BNC-2090, National Instrument).  The image was collected on a charge-coupled digital camera (CCD, LCL-902C, Watec) with total magnification of 100X. To ensure thermal stability, most of the microscope (except for the observation ports) as well as the syringe pumps were enclosed by a home-made acrylic microscope enclosure (28" $\times$ 25" $\times$ 18") heated with a heater fan (HGL419, Omega), and the temperature was maintained at 32°C by a proportional integral derivative temperature controller (the "incubator" mentioned above; CSC32J, Omega). A uniform temperature profile inside the incubator is achieved by active air flow from two small-profile heaters inside the enclosure.

DIC images of multiple fields-of-view were recorded at 2 frames/min and the focus was adjusted automatically by a total-internal-reflection (TIR) based autofocusing control loop. The back-reflected beam of a TIR-aligned 633 nm laser (LHRP-0081, Research Electro-Optics) impinges on a quadrant photodiode detector. The amplified difference signal is the error signal that is used as a feedback for adjustment of the objective (motor) position. A Virtual Instrument routine (LabView 7.0, National Instrument) was used to control all the components (i.e., sample stage, autofocus, pumps, CCD, and LED) and run the experiment for extended (>20 hrs) periods of time.

**Data analysis and construction of population growth curves.** The stack of acquired DIC images was loaded into ImageJ (NIH) and the division events of individual cells were tracked manually and recorded by a home-made plug-in. Cells that grew into long filaments or stopped reproduction were excluded from the analysis. The division event data was imported into Matlab (MathWorks) and processed. A typical periodic perturbation data set contained >200 cells. Note that only the original set of stalked cells was used for the present analysis; we did not include cell division data from any of the progeny cells that adhered to the glass surface.

Growth curves were constructed from the division event data for individual cells. Each observed division contributes to a unit increase in the population size. Since we only followed continuous reproduction of the initial population of stalked cells on the glass surface, the total number of cells generated by this population of cells can be represented by the differential equation $dN/dt = N_0/T_{div}$, where $N(t)$ is the number of total number of cells as a function of time, $N_0 = N(0)$, and $T_{div}$ is the division time of cells. Therefore, linear growth behavior is expected for the asynchronous division of a population of cells.

**Construction of phase difference distribution and characterization of synchrony.** The phase difference is defined as the temporal difference between the time of each cell division event (i.e., each dot in Fig. 2B) and the start time of the subsequent high inducer pulse (depicted in Fig. S3a inset). Without dispersion in the population, a single phase difference is present for a phase locking condition [32]. However, the existence of noise leads to a distribution of phases for the population under phase locking. To construct this distribution, we included division events between the start times of two successive high inducer pulses for periodic perturbations with an external period longer than 65 min; and for external periods shorter or equal to 65 min, we included division events between the end times of two successive high inducer pulses.

The synchronization index [31] is calculated based on the phase difference distribution, which is based on the Shannon Entropy: $SI = (S_{max} - SE)/S_{max}$, where $S_{max} = \ln N$, $SE = -\sum_{i=1}^{N} p_i \ln p_i$, $p_i$ is the probability at each state and N is the number of bins. SI ranges from 0 (uniform phase distribution) to 1 (singular phase occupation). In all of our calculation, we use a constant bin number of N = 20. The evolution of phase locking is characterized by the arithmetic mean of the phase difference distribution at different multiples of the external period (Fig. 2 and Fig. S5).

Construction of experimental phase response curves from cell division times. With the assumption that phase responses of individual cell cycle oscillators are independent of the pulse number (which is validated in the Supplementary Information), we constructed the phase response curve from the data of all pulses in each constant pulse period experiment. The scattered data (Fig. S3B) within a chosen bin range (i.e., 2 min as represented by the gray bar) were used to construct perturbed cell cycle time distributions (insets). For the distributions which are obviously truncated due to data sampling limitations, we used the center of the fitted Gaussian as the perturbed cell cycle time; while for other distributions, the arithmetic mean values are used instead as the reset cell cycle time.

By this approach, we obtained a set of phase response data (i.e. phase vs. perturbed cell cycle time) for each $t_H$. The results for $t_H = 10$ min and 15 min are shown in Fig. 3B and Fig. S2, where the perturbed cell cycle time is converted to phase advance or delay and the phase difference in minutes is scaled to be between 0 to 1 by the mean native cell cycle time at low inducer concentration (68.1 min). The missing data points for phase approaching unity are due to the finite width of high inducer concentration pulse. The phase response data at the minimum phase (i.e., phase = 0.011) are duplicated to indicate the periodic nature of phase response curves (i.e., these data are duplicated at phase = 1.011). These data are then fit with a trigonometric polynomial of degree three to ensure periodicity:

$y = a_1 \sin(2\pi x) + a_2 \sin(4\pi x) + a_3 \sin(6\pi x) + b_1 \cos(2\pi x) + b_2 \cos(4\pi x) + b_3 \cos(6\pi x) + c$. The fit parameters (a1, a2, a3, b1, b2, b3, c) for phase response curves at $t_H = 10$ and 15 min are (5.58725, 1.05375, - 0.03266, 1.585, 1.29886, 0.69753, 0.72275) and (8.41916, 1.60012, -0.32122, 1.7194, 2.57242, 1.06958, - 0.87352), respectively.

Assumptions underlying determination of the coupling function. We derive the form of Eq. (1) in the Supplementary Information, starting from a classic mathematical description of interacting oscillators [53]. We estimate the sensitivity function $Z(\varphi_1)$ from the gene regulatory network of the CtrA module [20]; more precisely, in numerical practice, we approximate $Z(\varphi_1)$ as constant over the duration of the pulse, with its value given by the published function at the phase when the pulse begins. The specific choice of the model in [20] does not significantly affect the result. In determining the coupling function $C_2(\varphi_1 - \varphi_2)$ as described in the Results and Discussion, we first analyze the steady-state solution in the case when there is a single pulse by assuming that: (1) $\omega_1$ and $\omega_2$ are equal to each other and (2) the effect of a pulse on the first oscillator is equally distributed throughout its duration. If the response of the first oscillator to the pulse is small and the steady-state is stable, the second oscillator will maintain the initial phase difference $\delta_0$. If the response of the first oscillator to the pulse is large, the second oscillator can be displaced to a new stable solution $\varphi_1(\infty) - \varphi_2(\infty) = \delta_i$ for $i \neq 0$. The criterion for a solution to be stable is $dC_2(\varphi_1 - \varphi_2)/d(\varphi_1 - \varphi_2)\big|_{\varphi_1 - \varphi_2 = \delta_i} > 0$. Because we do not observe discontinuous responses in the experiments, we conclude that the response of the first oscillator is not strong enough to allow solutions other than $\delta_0$. Rather than being instantaneous, the time required for the second oscillator to relax back to the initial state $\varphi_1(\infty) - \varphi_2(\infty) = \delta_0$ is set by the slope, with steeper slopes leading to faster relaxation.

Figure Captions

Fig. 1. Schematic for phase locking the stalked C. crescentus cell division cycle by periodically perturbing ctrA expression. (A) C. crescentus stalked cell cycle is driven by oscillating concentration of the master regulator protein, CtrA. The cell cycle begins with low CtrA concentration, allowing initiation of chromosome replication. CtrA levels then rise gradually, accompanied by cell growth and division. Cytoplasmic compartmentalization at the pre-divisional stage triggers the rapid proteolysis of CtrA, initiating another round of stalked cell division. (B) Schematic of phase locking. (Left) The expression of exogenous ctrA (in a mutant lacking endogenous ctrA) is controlled by a periodic inducer pulse train which oscillates between two discrete levels (Low and High), which then phase locks the dividing stalked cells on the surface of a microfludic flow channel (lower micrograph) as schematized on the right.

Fig. 2. C. crescentus cells can be phase locked. (A) Phase locking a population of single cells. The upper panel shows the cell growth trajectory overlapped with the external inducer pulse train (SI Text). The inserts are magnified views from 0 to 260 min and from 520 to 780 min. The lower panel shows the divisions of single cells (261 cells at pulse start) that were monitored for over 20 hrs. The timing of division events for individual cells are plotted (black dots) along lines parallel to the Time axis. Cells are arranged along the vertical axis according to their phases prior to the first perturbation (i.e., the diagonal line immediately before time zero). Inducer profile along experimental time is indicated in red, where high and low xylose levels are 0.03% (w/v) and 0.00009% (w/v) respectively. (B) Phase difference distribution. Phase difference (in minutes) between the internal cell cycle oscillator and external oscillator is analyzed. The distributions of phase difference after $2^{nd}$ and $10^{th}$ pulses are shown. (C) Quantification of phase evolution and division synchrony. The distributions in (b) are used to quantify the mean phase difference and synchronization. Both quantities are plotted with respect to the number of pulses delivered.

Fig. 3. The phase is more readily delayed than advanced. (A) Quantification of synchronization index under various external pulse profiles. The synchronization index ranges from zero to one as the population varies from asynchronous to synchronous. The synchronization indices (less the initial value) from the eighth to twelfth pulses are plotted for a variety of external periods ranging from 56 min to 89 min (converted to frequency) with 10 min and 15 min pulses (left vertical axis). The horizontal bars (right vertical axis) indicate the range for 1:1 phase locking of a noise-free cell cycle oscillator (Arnold Tongues). Such frequency ranges are inferred from the phase resetting curves in (a) and (b). $f_0$ is the intrinsic frequency. (B) Phase resetting curve (PRC) for 15 min pulses. The data (open circles) are fitted with a real trigonometric polynomial of degree three (solid line) to ensure periodicity. (C) Schematics for perturbations on CtrA oscillation by a single elevated ctrA expression pulse at two possible time points.

(D) Comparison between experimental and simulated 15-min-pulse PRCs based on the model of Li and Tyson[19].

Fig. 4. Proposed coupled-oscillator model for C. crescentus cell cycle control. (A) Interactions between core cell cycle regulatory module and cell division module. Cell division module is represented as looped connections of protein expression and interaction events. This closed loop is established by both protein interaction causalities and temporally connected events. The color scheme is the same as Fig. S9. The interactions are schematized in the lower panel. (B) Derived coupling function $C(\varphi_1 - \varphi_2)$. See Supplementary Information for details. (C) Reverse-calculated PRC overlapped with experimental PRC data for 15-min pulses. (D) Comparison between experimental PRC data for 10-min pulses and calculated PRC based on the coupling function in (B) derived from 15-min-pulse data. See SI Text for details.

Figures

Fig. 1.

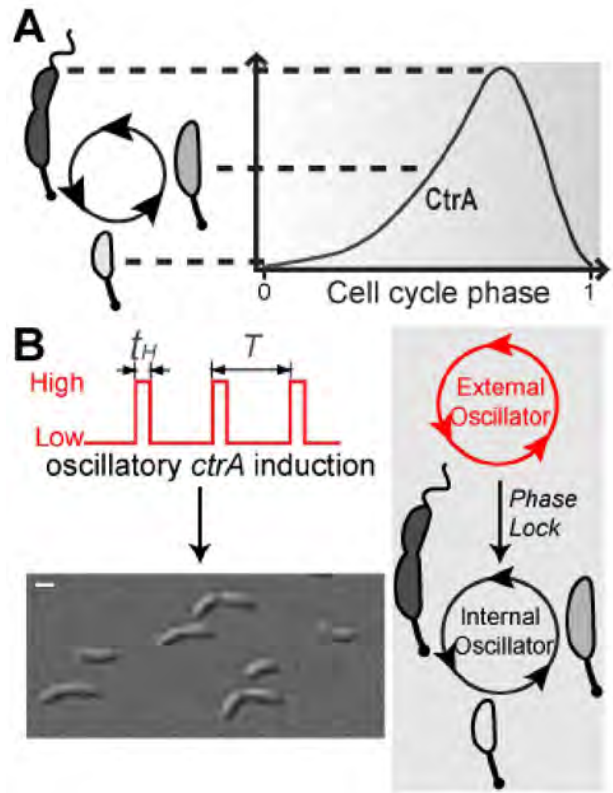

Fig. 2.

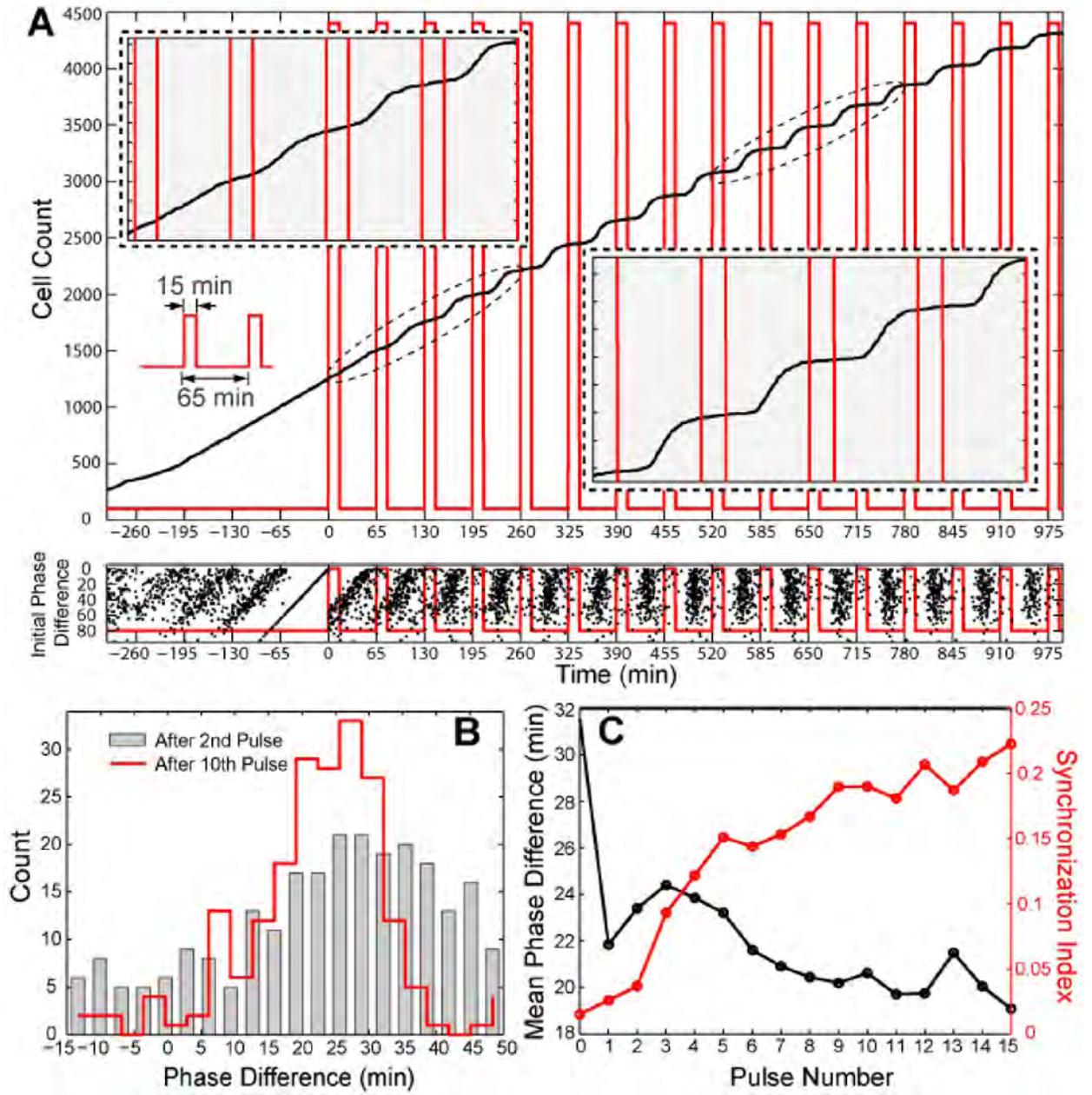

Fig. 3.

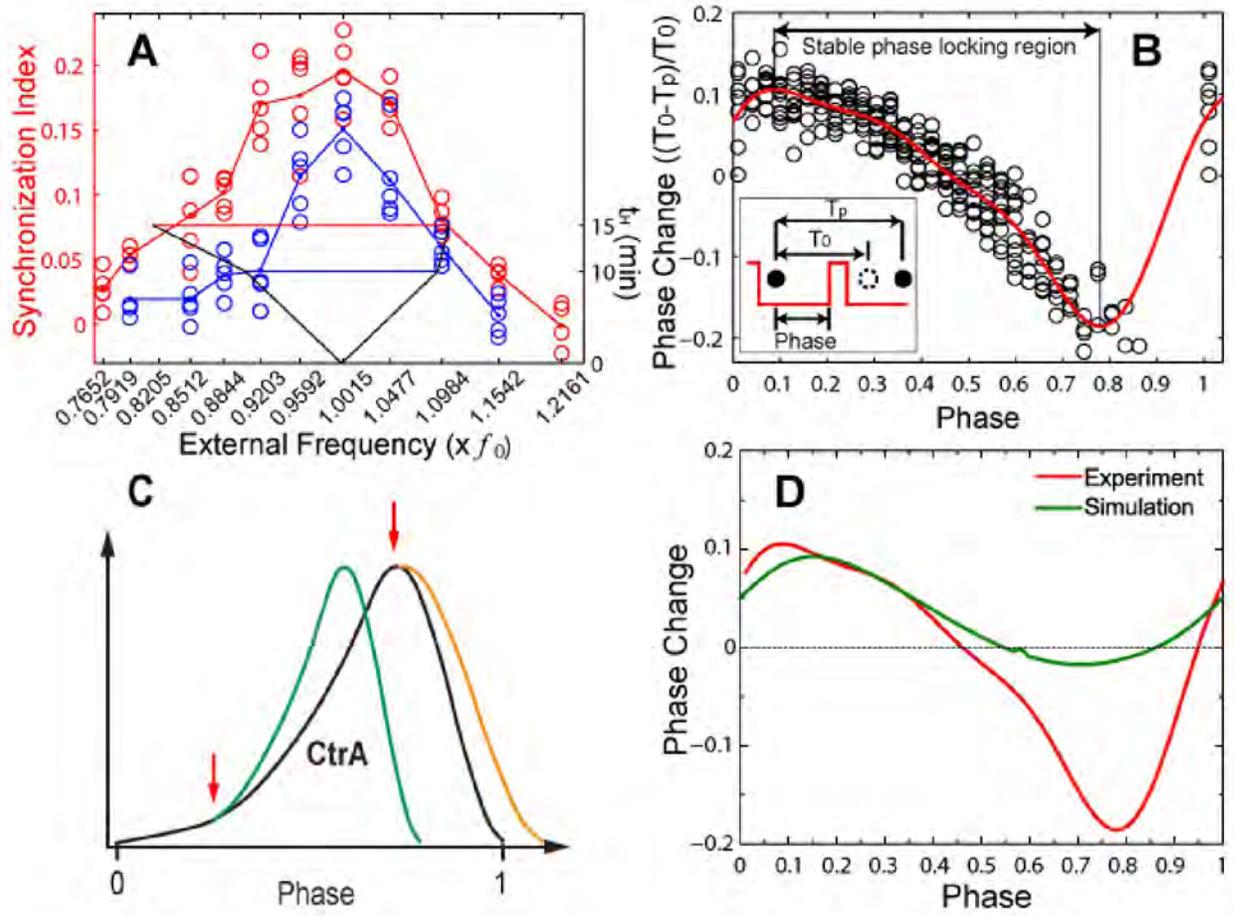

Fig. 4.

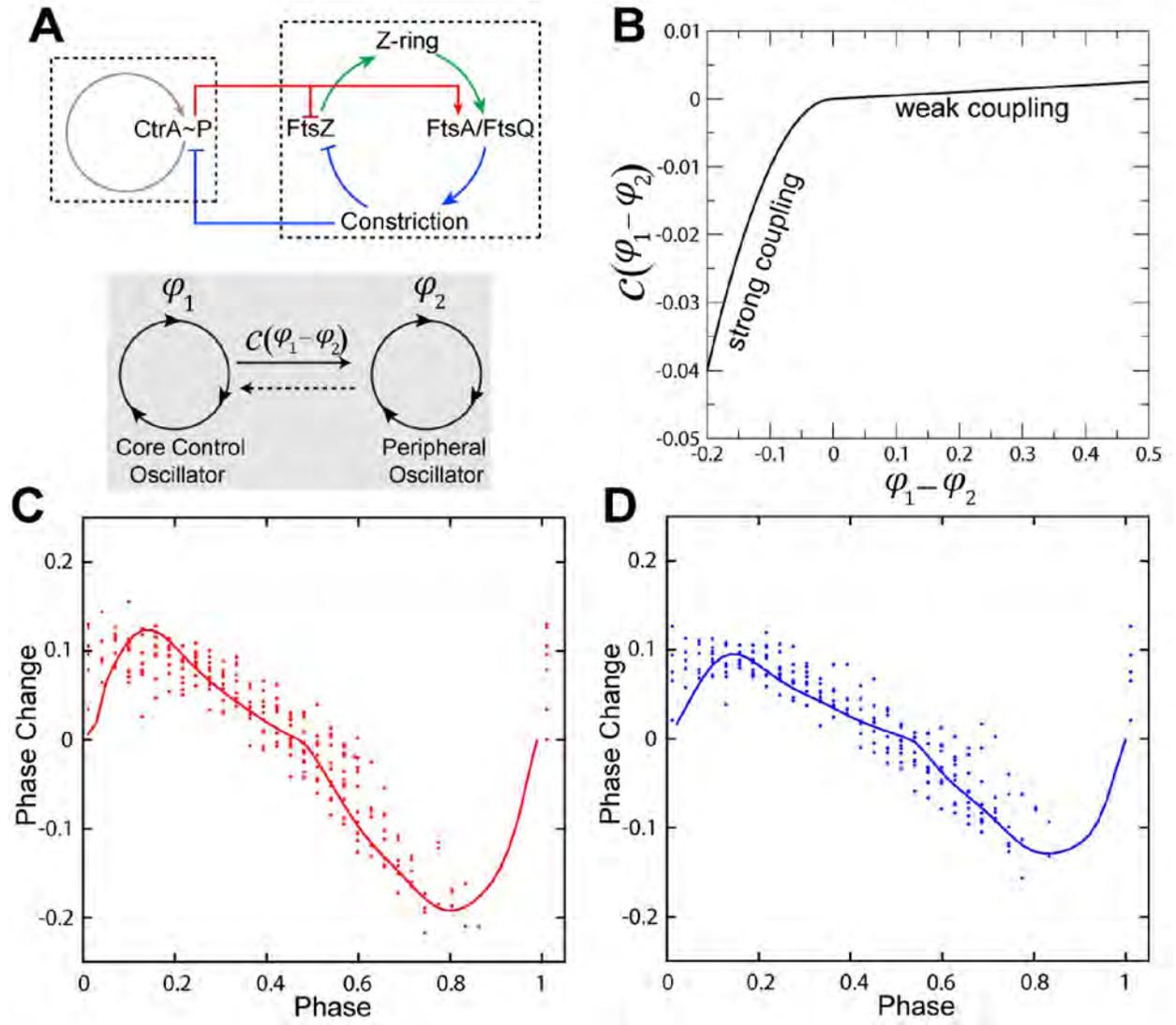

Phase resetting reveals network dynamics underlying a bacterial cell cycle

Yihan Lin[1, 4], Ying Li[2, 4], Sean Crosson[3], Aaron R. Dinner[1, 4, *], and Norbert F. Scherer[1, 4, *]

Department of Chemistry[1], Physics[2], and Biochemistry and Molecular Biology[3], and Institute for Biophysical Dynamics[4], University of Chicago, 929 East 57th Street, Chicago, IL 60637

*E-mail: dinner@uchicago.edu; nfschere@uchicago.edu

<u>Supplementary materials</u>

Contents

I. Assumption of fast relaxation of cell cycle oscillator after perturbation

Conventional oscillator perturbation experiments are based on isolated perturbations [1]. The phase response from such experiments is constructed by measuring the phase change stimulated by a single pulsed perturbation after sufficient relaxation of the oscillator back to its free-running mode (a limit cycle). In contrast, our analysis of phase response is based on periodic perturbation and assumes relatively fast relaxation to the steady-state after perturbation. To validate this assumption, we constructed the phase response curves from single cell measurements for each (different) external pulse period in a single periodic perturbation experiment ($t_H$ = 15 min and T = 80 min) (Fig. S3a). Qualitatively, the data from the different periods of the chemical perturbation lie within the same cell cycle dispersion envelope (i.e. along the Y-axis) and do not differ from each other in terms of dispersion. Moreover, the data from different low inducer durations (i.e., different relaxation times after perturbation) overlap well (Fig. S3b) except for the region where the sampling range of the cell cycle time is truncated due to construction (as shown in the insets). Therefore, it is valid to assume that the phase response curve obtained from our periodic perturbation experiments agrees with that from single pulse perturbation experiments.

II. From PRC to phase-locking: single-oscillator description and Poincaré return map analysis

By definition, a stable (amplitude) oscillator returns to its limit-cycle attractor after exposure to small amplitude perturbations. Regardless of the detailed molecular mechanism, the dynamics of the cell cycle oscillator in the presence of zero or small amplitude perturbations can be represented by its phase variable [2],

$$\mathrm{d}\varphi / \mathrm{dt} = \omega_0 + \varepsilon(t)Z(\varphi) + F(t) \tag{S1}$$

where $\omega_0$ is the intrinsic phase velocity, $Z(\varphi)$ is the sensitivity function, $\varepsilon(t)$ is a small external force that is applied as a function of time, and $F(t)$ is the noise. Integration of $Z(\varphi)$ over $\Delta t$, the duration of a single square wave perturbation, gives the phase resetting curve [2]:

$$\mathrm{PRC}(\varphi) = \int_{\varphi_0}^{\varphi(t+\Delta t)} \varepsilon(t)Z(\varphi)\mathrm{d}\varphi . \tag{S2}$$

For a periodic train of square pulses with frequency $f_{ext}$, which interacts with and entrains the cell cycle (i.e., the internal) oscillator about its fundamental frequency (termed 1:1 mode), we can characterize the phase-locking region and stability with Poincaré return map analysis [3]. The phase of the oscillator in the nth external period, $\varphi_n$, and its phase after another external period, $\varphi_{n+1}$, need to satisfy:

$$\varphi_{n+1} = [\varphi_n + \text{PRC}(\varphi_n) + \omega_0 \text{T}_{\text{ext}}] \bmod 1, \qquad (S3)$$

where $\omega_0$ is the intrinsic phase velocity, $\text{T}_{\text{ext}}$ is the external oscillator period, and $\varphi \in [0,1]$. Under phase-locking conditions, $\varphi_{n+1} = \varphi_n$ and we have

$$\varphi_n = \varphi_n + \text{PRC}(\varphi_n) + \omega_0 \text{T}_{\text{ext}} - 1 \qquad (S4)$$

or

$$\text{PRC}(\varphi) = 1 - \omega_0 \text{T}_{\text{ext}}, \qquad (S5)$$

which defines the range of external oscillator frequency that permit phase-locking. The range of internal oscillator phase values that permit phase-locking can in turn be obtained from the analysis of the Poincaré return map of Eqn. (S5). Two fixed points are found (Fig. S4) and their stabilities can be determined by the slope of the PRC [2,3]. Specifically, for the fixed point to be attractive and stable, it requires:

$$| f(\varphi_n)^{\cdot} | = \left| \frac{\text{d}[\varphi_n + \text{PRC}(\varphi_n) + \omega_0 \text{T}_{\text{ext}} - 1]}{\text{d}\varphi_n} \right| < 1,$$

or:

$$-2 < \text{dPRC}(\varphi)/\text{d}\varphi < 0. \qquad (S6)$$

Therefore, stable phase locking can be achieved only within the region where the slope of the PRC is between -2 and 0 [3].

An independent approach to identifying the stable phase-locking region is to construct the relation between the stable phase difference and the external period according to Eqn. (S5) by using the identified stable phase difference measured for various phase-locking experiments as in Fig. S5a. Specifically, the stable phase difference in time is converted to phase (between 0 and 1) as the horizontal axis. The difference between external period and internal unperturbed period (68.1 min) is converted as phase change (i.e., advance or delay) for the vertical axis. The constructed relations are overlaid on the PRCs as shown in Fig. S5b (open circles).

III. Numerical simulation of responses of noisy single phase oscillators to periodic and non-periodic perturbation

To examine whether a single oscillator model with Gaussian noise can capture the experimental results, we simulated the trajectory of a single phase oscillator using Eqn. (S1) assuming Gaussian white noise in the simulation. We assumed that the phase in the sensitivity function $Z(\varphi)$ increases linearly with respect to time during the high inducer pulse, hence Eqn. (S2) becomes:

$$\text{PRC}(\varphi) \equiv \text{PRC}(\varphi_0) = \int_{\varphi_0}^{\varphi_0 + t_H / T_0} Z(\varphi) d\varphi , \qquad (S7)$$

where $t_H$ is the high inducer pulse duration and $T_0$ is the mean native cell cycle time. With such simplification, we were able to approximate the analytical form of $Z(\varphi)$. The purpose is to perform direct simulation of the differential equation Eqn. (S1).

The simulation of Eqn. (S1) was carried out using the method described in [4]. Briefly, a small increment of $\varphi$ is calculated by

$$\varphi(t + \Delta t) - \varphi(t) = \omega_0 \Delta + \varepsilon(t) Z(\varphi) \Delta + D\gamma \sqrt{\Delta t} , \qquad (S8)$$

where $\Delta t$ is a very small time interval (0.01 min), $\omega_0$ is the intrinsic phase velocity ($\omega_0^{-1} = 68.1$ min), $\varepsilon(t)$ is a periodic or non-periodic function ($\varepsilon(t) = 1$ during high inducer pulse and 0 during low inducer pulse), $D$ is the noise amplitude approximated from experimental distribution (0.018), and $\gamma$ is a Gaussian random number of zero mean and variance equals to 1. As above, $Z(\varphi)$ is the sensitivity function; it is important to note that, in applying Eqn. (S8), we assume that $Z(\varphi)$ is constant over the entire pulse duration, with the value specified by the phase at the time of initiation of the pulse.

A representative simulation with a population of phase oscillators starting with randomly distributed phase is shown in Fig. S6, which captures the phase locking feature of its experimental counterpart (Fig. 2A). Simulations conducted for different pulse profiles were used to quantify the synchronization index and phase difference, as have been characterized for the experiments. The trends of both quantities (Fig. S7a and Fig. S7 b-c) agree with the experimental results (Fig. 3a and Fig. S5 b-c). Furthermore, simulations with non-periodic external pulse perturbations also capture the behaviors of this cell cycle oscillator measured under non-periodic perturbation experiment (Fig. S8). It is noted that these simulations only complement the analysis done in the previous section and do not offer new insights.

IV.     Comparison between experimental and simulated PRCs

A key test was to establish whether the published (molecularly detailed) model of the network could capture the experimental PRC. Using a model of the Caulobacter cell cycle from Li and Tyson [5], we simulated the CtrA perturbation experiment and used the response of the model to construct the PRC. This model incorporates details pertinent to the regulation of the cell cycle including known regulatory proteins oscillations and subcellular processes such as DNA replication and Z-ring formation. This model captures the phenotypes of various existing mutants and provides predictions for novel mutants. The  ctrA / $P_{xyl}$-ctrA mutant is relevant to our experiment. We adopt most of the model parameters and simulate a single-pulse perturbation experiment to calculate the PRC (Fig. S9a). Since the unperturbed cell cycle time in the simulation of this mutant is longer than our experimental value (97.1 min vs. 68.1 min), the duration of the simulated pulsed perturbation with elevated ctrA induction is scaled accordingly (i.e. 14.3 min and 24.1 min in simulation vs. 10 min and 15 min in experiment). The perturbation amplitude (i.e., the elevated ctrA induction rate) was adjusted such that the relaxation from the perturbation is essentially complete within one cell cycle and the magnitude of the phase advance in the simulated PRCs is close to that observed experimentally.

The simulated PRCs exhibit a sinusoidal shape that is similar to the experimental ones. However, the experimental PRCs exhibit much larger phase delay responses when perturbation is administrated during the latter portion of the cell cycle (Fig. S9a). Since the phase response characterizes the response of the downstream cell division machinery to the perturbation of CtrA concentration, we need to examine by simulation how the perturbation of CtrA concentration propagates through the entire network to the regulation of cell division machinery. Thus, we summarize the major effects of CtrA on various cell division proteins and relevant division processes, including the production of FtsZ, the formation of Z-ring by FtsZ, the production of FtsQ, and the constriction of Z-ring by FtsQ (Fig. S9b). In the unperturbed cell cycle, the division process starts with the expression of FtsZ in the early stalked cell [6]. The increase in FtsZ leads to formation of Z-ring in the mid-cell plane. As CtrA~P increases rapidly in the late stalked stage, the expression of FtsZ is suppressed while that of FtsQ (and FtsA) is activated [6,7]. The fully-assembled Z-ring begins to be constricted by FtsQ (and FtsA). As constriction progresses, the stabilities of FtsZ [6,8] and FtsQ drop. Turnover of FtsZ and FtsQ accompany the completion of constriction (i.e., cell division). The simulated protein trajectories (solid) in Fig. S9b reflect the aforementioned regulations. For $t_H$ = 24.1 min simulation, when perturbation on CtrA expression is introduced early in the cell cycle phase (i.e., phase = 0.2), cell division is advanced significantly with a comparable magnitude as the advance in CtrA~P oscillation: the FtsQ peak appears earlier and constriction of Z-ring completes faster, and the magnitude of advance of FtsQ or Z-ring trajectory is comparable to that of CtrA~P (left, Fig. S9b). However, the delay response of cell division to CtrA perturbation later in the cell cycle (i.e., phase = 0.75) is much less significant (right, Fig. S9b) than previous situation (left, Fig. S9b). These molecular-level details of perturbation-response from simulations underlie the shape of the simulated PRC (right, Fig. S9a). The regulation of CtrA on the division process in the model simulation, i.e., the advance/delay in CtrA oscillation, propagates almost linearly to the advance/delay in the division process.

With these mechanistic insights, we now try to resolve the discrepancy between experimental and simulated PRCs by parameter tuning and limit our focus on the longer $t_H$ duration PRCs (right, Fig. S9a). Since the perturbation of CtrA is constant, it should be the interaction between CtrA and the cell division process that contributes to the observed discrepancy. And since much of the discrepancy lies in the delayed portion of the PRCs, we need to tune the relevant parameters that would cause larger delay in the cell division with the same CtrA perturbation. Of the options, (i) strengthening the repression of FtsZ by CtrA, (ii) weakening the promotion of FtsQ by CtrA, and (iii) weakening the promotion of Z-ring constriction by FtsQ, only the first option can be realized in the model. Tuning (ii) and (iii) lead to instability in the simulation (left, Fig. S9c). However, even a ~30% decrease of the CtrA-FtsZ repression constant leaves the PRC almost unchanged (right, Fig. S9c). Thus, this inability to achieve the measured PRC by adjusting parameters of the model suggested that the discrepancy was more fundamental in nature.

## V.    Derivation of coupled-oscillator model

As discussed in the main text, there is evidence that cell cycle modules oscillate autonomously and we interpret measurements with a coupled oscillator model. Based on classic mathematical descriptions of interacting oscillators [9], we start from the ordinary differential equations for the concentration of proteins ( $\vec{X}_1$ and $\vec{X}_2$ ) in these two modules:

$$\frac{d\vec{X}_1}{dt} = \vec{F}_1(\vec{X}_1) + \varepsilon \vec{V}_1(\vec{X}_1, \vec{X}_2)$$

$$\frac{d\vec{X}_2}{dt} = \vec{F}_2(\vec{X}_2) + \varepsilon \vec{V}_2(\vec{X}_1, \vec{X}_2),$$

$$(S9)$$

where vectors $\varepsilon \overset{1}{V}_1(\overset{1}{X}_1, \overset{1}{X}_2)$ and $\varepsilon \overset{1}{V}_2(\overset{1}{X}_1, \overset{1}{X}_2)$ weakly couple the two oscillators ($\varepsilon$ is small). In addition, we assume that the frequencies of the two oscillators are close to each other when there is no coupling. To emphasize this feature, we rewrite $\overset{1}{F}_i$ as

$$\overset{1}{F}_i(\overset{1}{X}_i) = \overset{1}{F}(\overset{1}{X}_i) + \varepsilon \overset{1}{f}_i(\overset{1}{X}_i) \qquad (S10)$$

To apply this formalism to the present case, we need to transform the concentration representation into a phase representation. We denote the phases of the two oscillators as $\varphi_1(\overset{1}{X}_1)$ and $\varphi_2(\overset{1}{X}_2)$. We choose the units of time such that $\dfrac{d\varphi_1}{dt} = 1$. Comparison with Eqn. (S9) with $\varepsilon = 0$ e $= 0$ and the chain rule then suggest

$$\overset{1}{F}(\overset{1}{X}_i)\nabla_{\overset{r}{X}_i}\varphi(\overset{1}{X}_i) = 1. \qquad (S11)$$

Below, we denote $\nabla_{\overset{r}{X}_i}\varphi(\overset{1}{X}_i)$ as $\overset{1}{Z}(\varphi_1)$ to reflect its role as the phase response function. In this way, the equations of motion of the phases are

$$\frac{d\varphi_1}{dt} = 1 + \varepsilon[\overset{r}{Z}(\varphi_1)\overset{r}{V}_1(\varphi_1, \varphi_2) + \overset{r}{Z}(\varphi_1)\overset{r}{f}_1(\varphi_1)]$$
$$\frac{d\varphi_2}{dt} = 1 + \varepsilon[\overset{r}{Z}(\varphi_2)\overset{r}{V}_2(\varphi_2, \varphi_1) + \overset{r}{Z}(\varphi_2)\overset{r}{f}_2(\varphi_2)] \qquad (S12)$$

where $\overset{1}{V}_i[\varphi_1(\overset{1}{X}_1), \varphi_2(\overset{1}{X}_2)] = \overset{1}{V}_i(\overset{1}{X}_1, \overset{1}{X}_2)$ and $\overset{1}{f}_i[\varphi_i(\overset{1}{X}_i)] = \overset{1}{f}_i(\overset{1}{X}_i)$.

As an approximation, we consider the influence of the coupling terms on the phases to be homogenous for all $\varphi_i$ values and integrate over the period T and $\varphi_i = t + \text{const}$ (small $\varepsilon$). Eqn. (S12) becomes:

$$\frac{d\varphi_1}{dt} = \omega_1 + C_1(\varphi_1 - \varphi_2)$$
$$\frac{d\varphi_2}{dt} = \omega_2 + C_2(\varphi_1 - \varphi_2) \qquad (S13)$$

where $\omega_i = 1 + \varepsilon \dfrac{\int_0^T \overset{r}{Z}(\varphi_i(t))\overset{r}{f}_i(\varphi_i(t))dt}{T}$ and $C_i(\varphi_1 - \varphi_2) = \varepsilon \dfrac{\int_0^T \overset{r}{Z}(\varphi_i(t))\overset{r}{V}_i(\varphi_1(t), \varphi_2(t))dt}{T}$.

Now, we consider the situation in which the ctrA induction rate follows a periodic square wave $\overset{1}{\Delta}(t)$:

$$\frac{d\vec{X}_1}{dt} = \vec{F}_1(\vec{X}_1) + \varepsilon\vec{V}_1(\vec{X}_1, \vec{X}_2) + \vec{\Delta}(t)$$

$$\frac{d\vec{X}_2}{dt} = \vec{F}_2(\vec{X}_2) + \varepsilon\vec{V}_2(\vec{X}_1, \vec{X}_2).$$

$(S14)$

Following the same transformation, the phase equations are

$$\frac{d\varphi_1}{dt} = \omega_1 + C_1(\varphi_1 - \varphi_2) + \vec{Z}(\varphi_1)\vec{\Delta}(t)$$

$$\frac{d\varphi_2}{dt} = \omega_2 + C_2(\varphi_1 - \varphi_2)$$

$(S15)$

where $\omega_1$ and $C_1(\varphi_1 - \varphi_2)$ are the same as above.

In the experiments, the readout is the phase response curve of the cell division oscillator in the presence of periodic ctrA induction. We ignore $C_1(\varphi_1 - \varphi_2)$ in the equation for the CtrA oscillator because we assume that the external driving force $\vec{\Delta}(t)$ is larger than the coupling $\varepsilon$. This is reasonable because the effect of the external driving on the second oscillator is of order $\varepsilon\Delta$ but the effect on the first oscillator is of order $\varepsilon^2\Delta$ because the perturbation must feed back from the second oscillator. These considerations yield Eqn. (1) in the main text.

Fig. S1. Cell cycle time distributions for single cells of ctrA mutants for constant xylose conditions. Division times of FC1006 are characterized for different constant xylose concentrations. Specifically, the statistics are $68.1 \pm 15.6$ min for 0.00009% xylose (mean $\pm$ SD, N=5160), $64.2 \pm 9.0$ min for 0.00027% xylose (N=444), and $65.5 \pm 9.2$ min under 0.03% xylose (N=244). And for reference, $65.3 \pm 9.9$ min for FC1071 for no xylose (N=311). The numbers of "superfast" division events (i.e. below or equal to 33min) are 14 out of 5160 for the 0.00009% condition, 2 out of 444 for the 0.00027% condition, and zeros for the other two.

Fig. S2. Phase resetting curve for $t_H =10$min. The data (open circles) are fitted with real trigonometric polynomial of degree three (solid line) to ensure periodicity.

Fig. S3. Fast relaxation of this cell cycle oscillator after perturbation. (a) Phase responses from different external pulse periods from a single periodic chemical perturbation experiment. The data are from the same periodic perturbation experiment ($t_H = 15$min and T = 80min). The responses (characterized as perturbed cell cycle time) for different perturbations are separated and plotted. The inset illustrates the construction of the x and y components in the plot. (b) Phase responses for different low inducer durations (i.e. relaxation time) at $t_H = 15$min: T = 80min and T = 59min. The responses from all pulse periods for each periodic condition are plotted. The inset in the upper left corner shows the distribution of perturbed cell cycle time for the 8⩽ phase <10 min at T = 59min and the inset in the lower right corner shows the distribution for 54⩽ phase <56 min at T = 80min. Both distributions are fitted with a Gaussian (red).

Fig. S4. Poincaré return map analysis and experimental construction of stable phase-locking region. Poincaré return map for Eqn. (S4). Dashed line represents $\varphi_{n+1} = \varphi_n$. Solid curve maps the initial phase onto itself by using PRC at $t_H =15$min at a certain $T_{ext}$ value. It intersects with diagonal line giving rise to a stable fixed point (solid) and an unstable fixed point (open).

Fig. S5. Phase differences between internal cell cycle oscillator and external oscillator at phase locking agree with phase response curves. (a) Phase evolution under various external oscillator profiles. The upper panel includes data for 10 min high xylose concentration pulse duration ($t_H = 10$ min) with external oscillator period ranging from 59 to 86 min. The bottom panel includes data for $t_H = 15$ min with external oscillator period ranging from 56 to 89 min.

(b) Phase difference between internal cell cycle oscillator and external oscillator at phase locking for a variety of external periods with $t_H = 10$min. The phase differences from 8[th] pulse to 12[th] pulse for the experimental conditions shown in (a) are overlapped with phase response curve (solid) from Fig. 3B and Fig. S2. (c) Analogous plot as (a) for $t_H = 15$min.

Fig. S6. Phase locking simulations with Eqn. (S1). Numerical simulation counterpart for experiment shown in Fig. 2A.

Fig. S7. Simulated phase locking dynamics capture experimental measurements. (a) Counterpart of Fig. 3A from simulation. (b-c) Counterparts of Fig. S5 b-c.

Fig. S8. Synchronization of single C. crescentus cells by non-periodic external perturbations with $t_H = 15$min. (a) Experiment trajectories measured for different non-periodic perturbation profiles. Top panel shows the perturbation with periods (i.e. T) as Gaussian random numbers with mean of 75 min and variance of 25 min$^2$. Middle panel shows the perturbation with a frequency "up-chirp", i.e. the period decreases by 2min in each successive interval. Bottom panel shows the perturbation with a frequency "down-chirp", i.e. the period increases by 2min successively. (b) Simulation counterparts for the experiments in (a).

Fig. S9. Discrepancies between our experimental PRCs and simulated PRCs from the Li-Tyson model [5]. (a) Comparison between experimental and simulated PRCs. The simulations are carried out based on the ΔctrA + constitutive $P_{xyl}$-ctrA mutant as described (Figure 8 of [5]) with k'$_{ctrA}$ modified to 0.03. This mutant has a native division time of 97.1min without perturbation and a finite width square-wave perturbation (k'$_{ctrA}$: 0.03 ≙ 0.05) is introduced at various cell cycle phases. The cell cycle phase change is quantified by the change in cell cycle time. Left: Perturbation simulation is done with $t_H = 14.3$min to compare with $t_H = 10.0$min experiment (i.e. $t_H$ occupies the same fraction of cell cycle time). Right: Perturbation simulation is done with $t_H = 21.4$min to compare with $t_H = 15.0$min experiment. (b) Molecular details of the perturbation-response simulations. (Top) Regulatory circuit for Z-Ring formation and constriction. Transcription of ftsZ starts early in the stalked cell cycle and the level of FtsZ protein, which assembles into the Z-Ring at the mid-cell plane, peaks at the late stalked stage while its transcription is repressed by CtrA~P. The regulators for Z-Ring constriction, FtsQ and FtsA, are transcribed from a common promoter $P_{QA}$ which is induced by CtrA~P at the late stalked stage. (Left) The perturbed and unperturbed trajectories of various protein species in the Li-Tyson model [5] that are subject to square-wave ctrA induction perturbation in the early cell cycle phase. The perturbation time window is indicated by the filled box. Unperturbed trajectories are shown as solid curves while the perturbed ones are in dashed. (Right) Same as (Left) for perturbation introduced at a later cell cycle phase. (c) Tuning of the network parameters fails to capture experimental PRC. (Left) Permissive and restrictive nodes for parameter tuning. The green box indicates the interaction (CtrAP ⊣ FtsZ) whose binding parameter (i.e. JiFtsZCtrA [5]) can be changed by >10%. The pink boxes indicate the opposite cases (JiFtsQCtrA for CtrAP → FtsAQ and JZFtsQ for FtsAQ → Constriction). (Right) The PRC from model simulation is not sensitive to the allowed range of parameter changes. The PRC obtained by lowering JiFtsZCtrA by ~30% (i.e. 0.7 à 0.5, maximum allowed change) is shown (blue) and compared with the PRC without parameter change (green, same as the one in the right panel of (a)).

Fig. S10. Phase resetting curves of the core regulatory module (CtrA) based on our previous model [10]. (Left) $t_H$ = 10 min. (Right) $t_H$ = 15 min.

Fig. S11. Schematic of transcription regulations and protein interactions for Z-ring assembly and constriction. CtrA~P interacts with division process by transcription regulation of ftsZ and ftsQA.

Fig. S12. Asymmetric coupling produces skewed cell cycle noise as observed in experiment. Eqn. (1) in the main text was simulated with noise to produce the cell cycle time distribution under unperturbed condition and to compare with experiment (blue here; gray in Fig. S1,). White noises ($\xi_1, \xi_2$) are attributed to the two oscillators, $\varphi_1$ and $\varphi_2$ respectively. The simulated distribution (green) is generated with $\langle \xi_1^2 \rangle = 0.001$ and $\langle \xi_2^2 \rangle = 0.0001$.

Fig. S1.

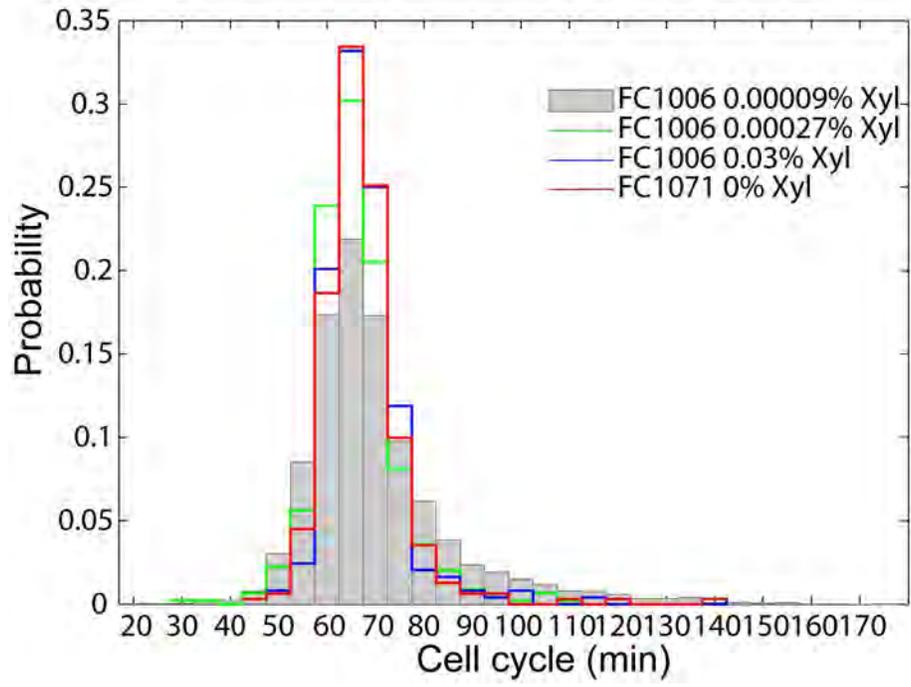

Fig. S2.

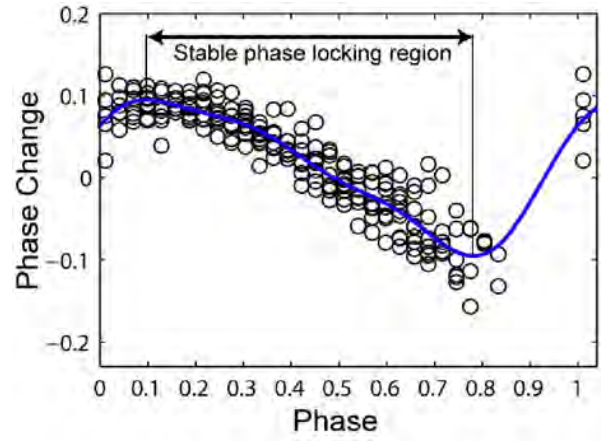

Fig. S3.

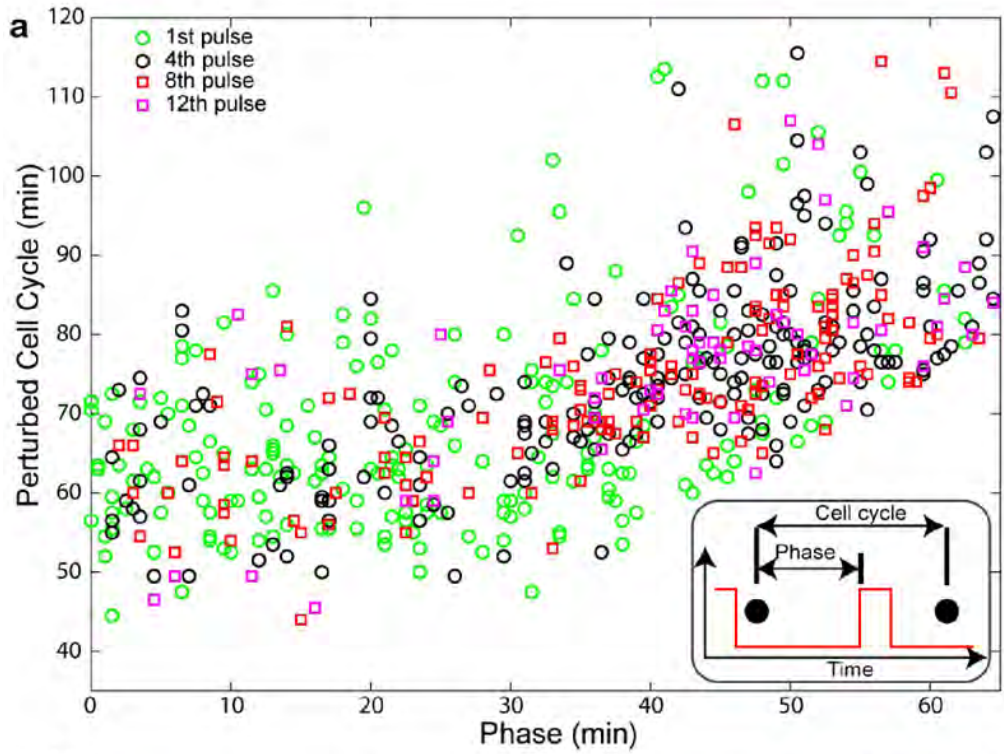

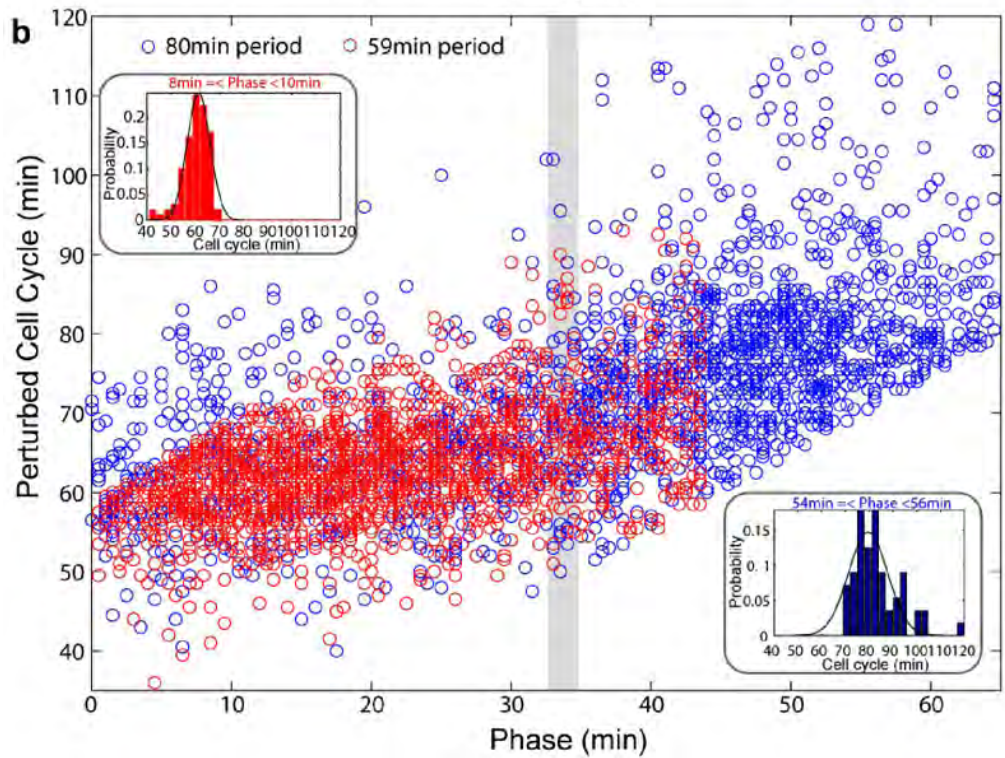

Fig. S4.

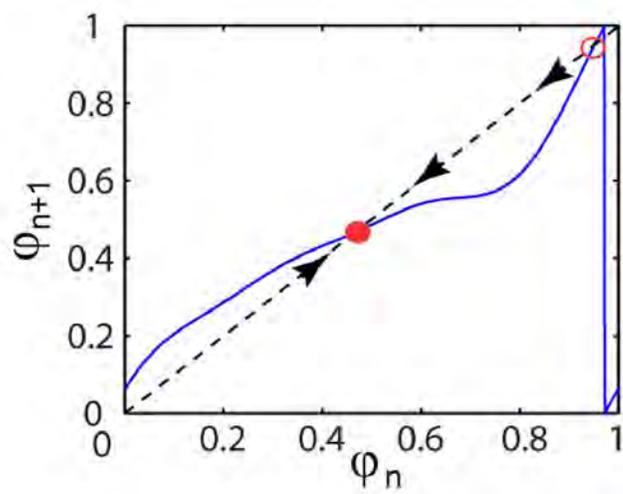

Fig. S5.

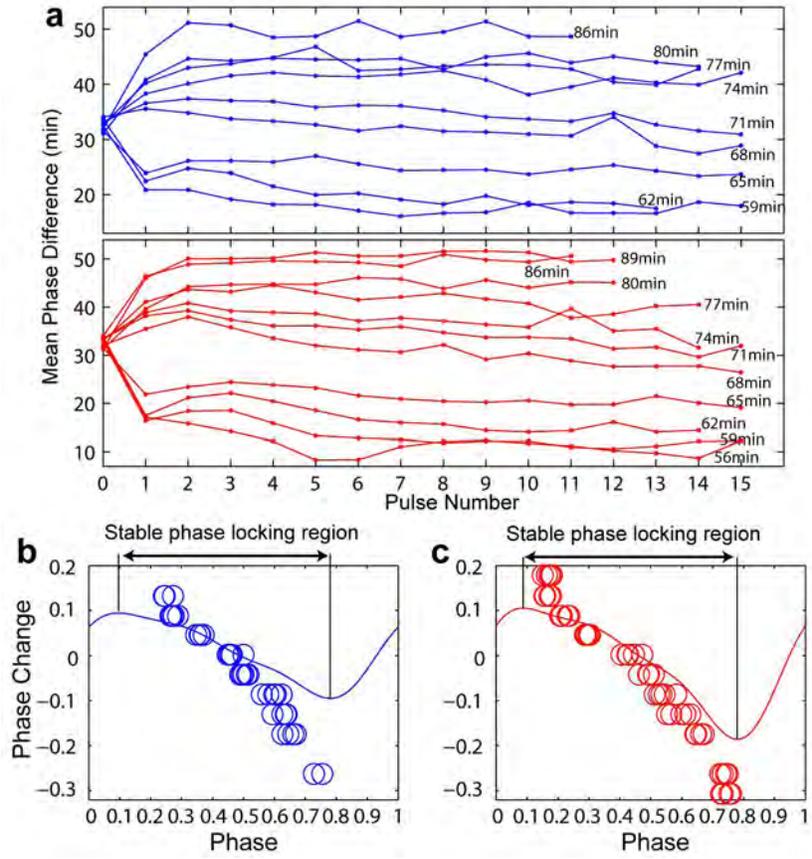

Fig. S6

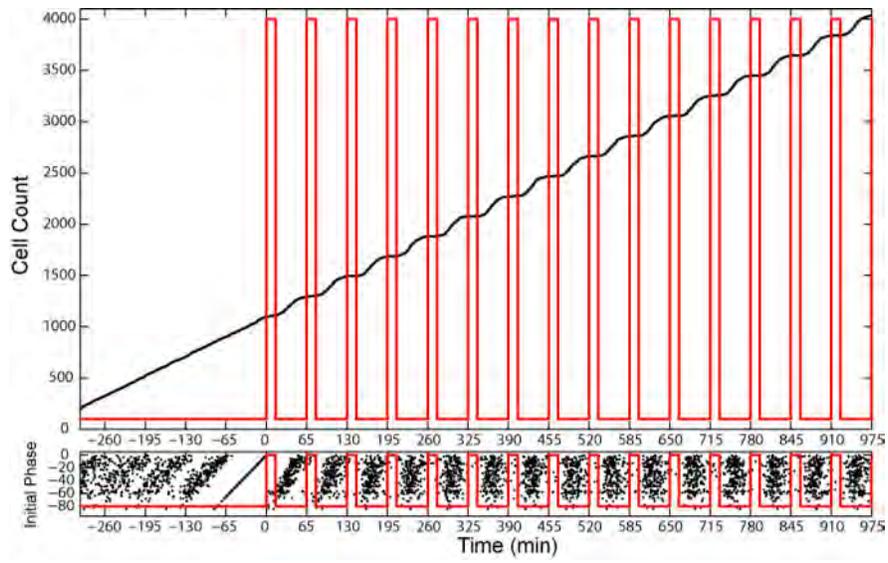

Fig. S7.

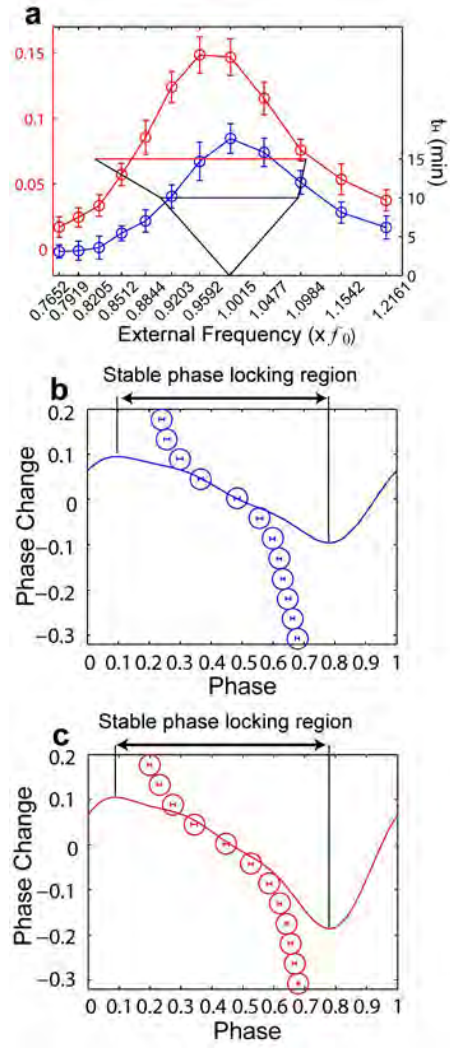

Fig. S8.

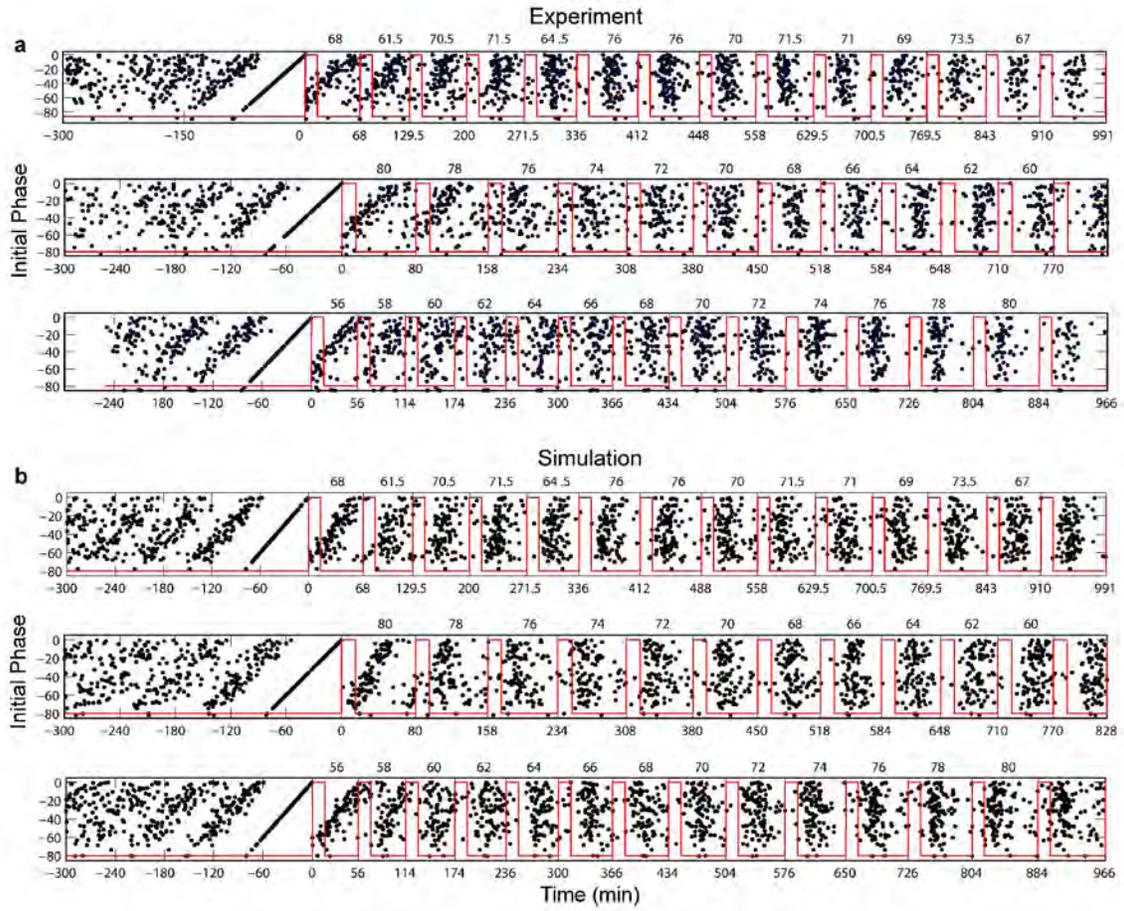

Fig. S9.

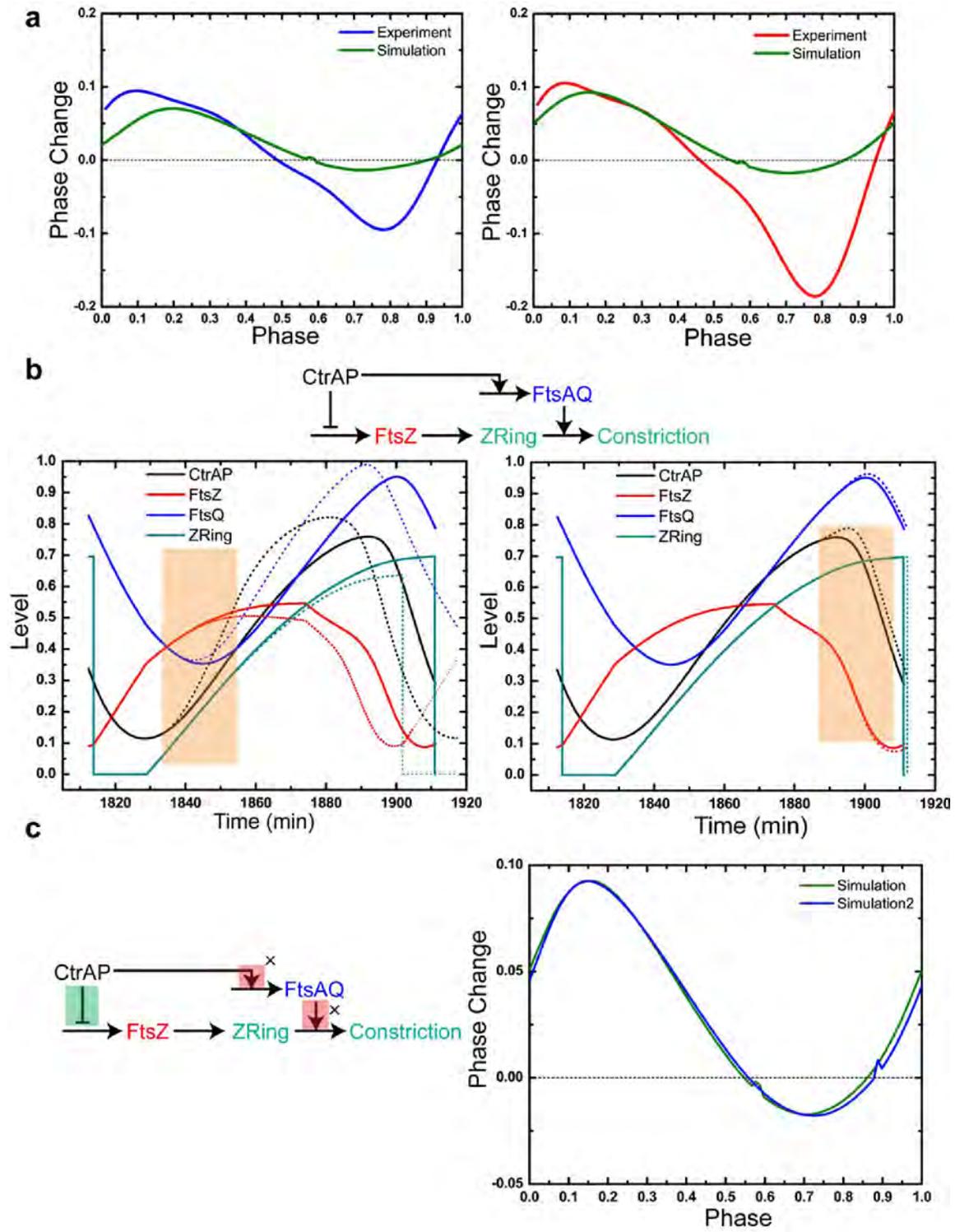

Fig. S10.

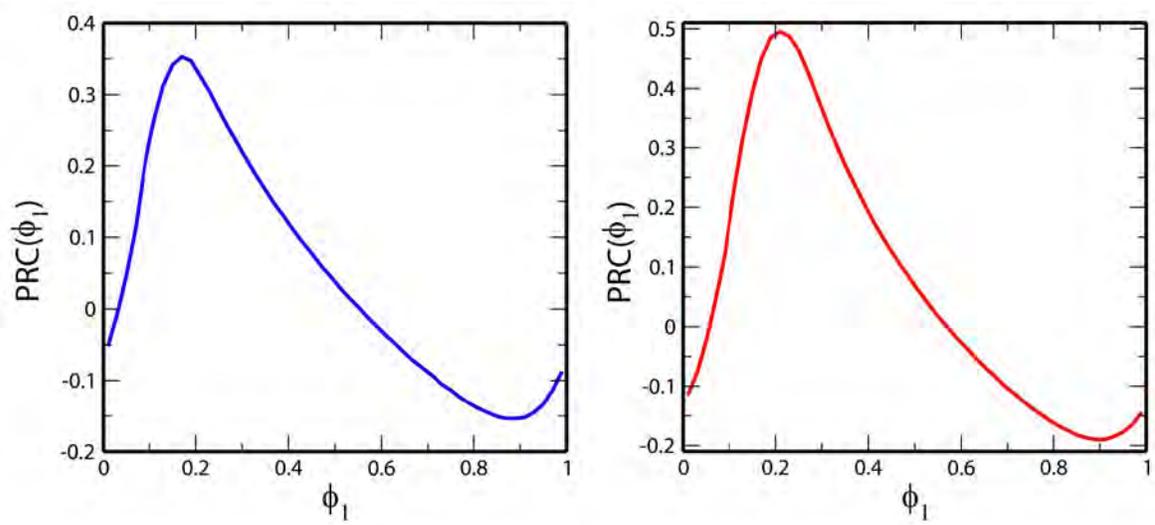

Fig. S11.

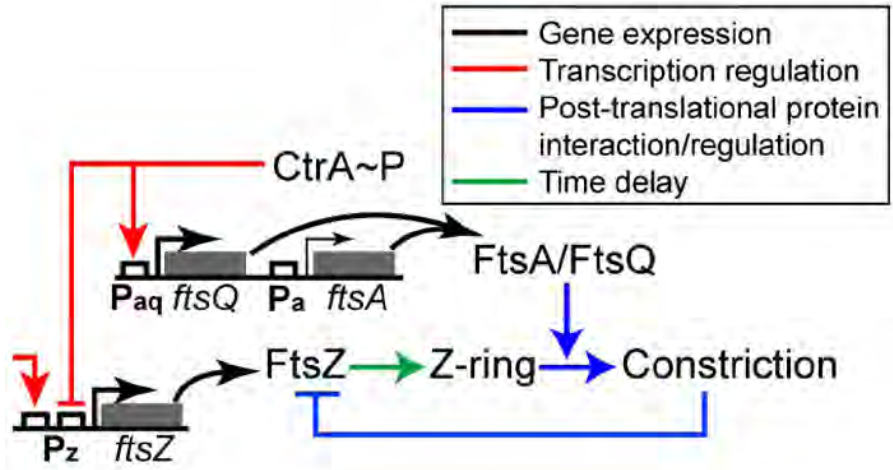

Fig. S12.

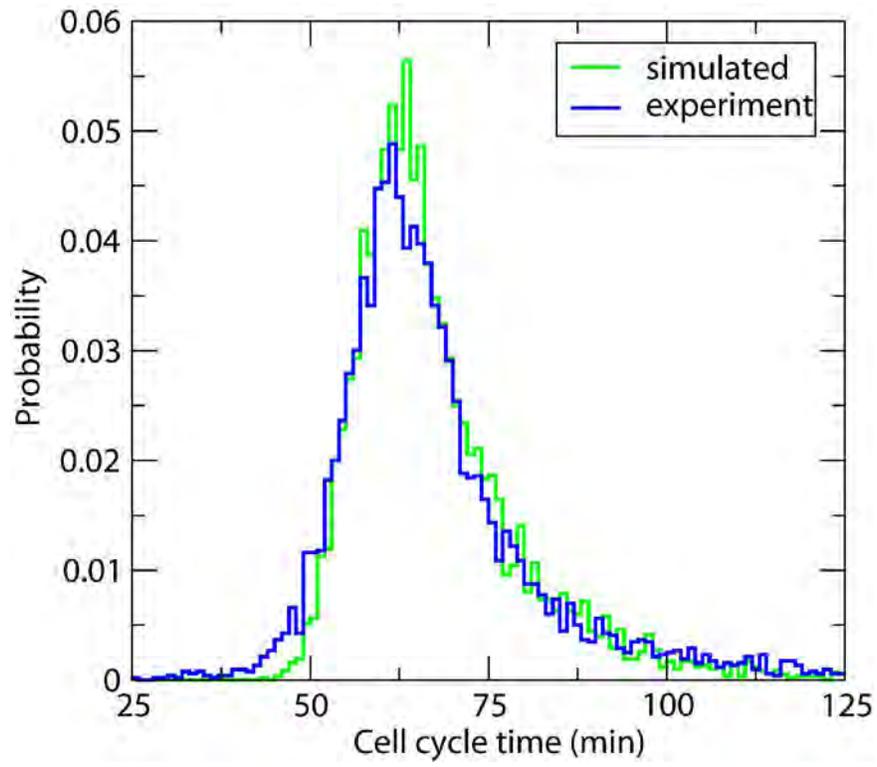